\input{aipcheck}

\documentclass[pdftex
    ,final            
  ]
  {aipproc}

\usepackage[rflt]{floatflt}

\layoutstyle{6x9}
\def\msun{M$_{\odot}$}
\def\rsun{R$_{\odot}$}

\def\sbs{SBS\,1150+599A}
\def\pn{PN\,G135.9+55.9}

\def\degs{\ifmmode ^{\circ}\else$^{\circ}$\fi}
\def\amin{\ifmmode ^{\prime}\else$^{\prime}$\fi}
\def\asec{\ifmmode ^{\prime\prime}\else$^{\prime\prime}$\fi}

\def\degs{\ifmmode ^{\circ}\else$^{\circ}$\fi}
\def\amin{\ifmmode ^{\prime}\else$^{\prime}$\fi}

\def\eqalign#1{\null\,\vcenter{\openup1\jot \m@th
   \ialign{\strut\hfil$\displaystyle{##}$&$\displaystyle{{}##}$\hfil
   \crcr#1\crcr}}\,}
\begin{document}

\title{Recycling Matter in the Universe. \\  {\small X-ray observations of SBS1150+599A (PN\,G135.9+55.9)}}


\author{Gagik  Tovmassian}{
  address={Institute of Astronomy, Universidad Nacional Autonoma de M\'exico }
}

\author{John Tomsick}{
  address={University California Berkeley, USA}
}

\author{Ralf Napiwotzki}{
  address={University of Hertfordshire, UK}
}
\author{Lev Yungelson}{
  address={Institute of Astronomy, RAS, Russia}
}

\author{Gra\.zyna  Stasi\'nska}{
  address={Observatoire du Meudon, France}
}

\author{Miriam Pe\~na}{
  address={Institute of Astronomy, Universidad Nacional Autonoma de M\'exico }
}
 
\author{Michael Richer}{
  address={Institute of Astronomy, Universidad Nacional Autonoma de M\'exico }
}

\begin{abstract}
We present X-ray observations of the close binary nucleus of the planetary nebula \sbs\  obtained with the XMM-{\sl Newton} satellite. Only one component of the binary can be observed in optical-UV. New X-ray observations show that the previously invisible component is a very hot compact star. This finding allows us to deduce rough values for the basic parameters of the binary. With a high probability  the total mass of the system exceeds Chandrasekhar limit and  makes the SBS1150+599A one of the best candidate for a supernova type Ia progenitor.   
\end{abstract}

\maketitle


\section{Introduction}
Supernovae  type Ia (SN~Ia)  are universally believed to be a result of an explosion  following the accumulation of a Chandrasekhar mass by a white dwarf (WD)  via merger of two lower-mass WDs or via  accretion of sufficient mass from a binary companion.  According to the first scenario  \cite{1984ApJS...54..335I,1984ApJ...277..355W},  components of the binary (double degenerate, DD)  merge due to the angular momentum loss through gravitational wave radiation. The alternative scenario \cite{1973ApJ...186.1007W} suggests that the companion is a non-degenerate star pouring  hydrogen-rich matter onto a WD via stellar wind or Roche lobe overflow.  So far, both scenarios  encounter difficulties in proving the route to SN Ia. The mechanism of accretion of several tenths of a \msun\ is not so evident. It is unclear whether there are enough symbiotic stars or semidetached systems to be progenitors of SN~Ia and how efficient the accretion is in these systems (see, e. g. \cite{2005ASSL..332..163Y} for discussion).  On the other hand, the SPY project revealed dozens of close DDs among more than thousand WDs surveyed  \cite{2004RMxAC..20..113N}, but there are only a few for which the total mass seems to exceed 1.4\msun\ (i.e., enough to  qualify them as  potential SN~Ia progenitors).

Here, we report on new observations of the planetary nebula \sbs\ (PN\,G135.9+55.9) that allow us to determine the composition of its close DD core and determine its place in the evolutionary path that probably leads to a type Ia SN event, completing a full  cycle in stellar evolution.

 \section{The revised parameters  based on new observations}

The new X-ray observations, which  came on heels of recently available HST data,  reveal  the component of the binary system invisible in the optical/UV range. When the binary nature of the object was discovered \cite{2004ApJ...616..485T}, it was naturally assumed that the observed hot WD-like object is the central star of planetary nebula (CSPN). Usually, this would imply that the residual core of the post-AGB star that ejected its envelope is the ionizing source for the nebula. Very blue (hot) continuum in optical was suggestive of that idea and  FUSE observations in far-UV range supported that too. The temperature  of CSPN was determined to be around 115\,000\,K based on the data and assumptions described in \cite{2004ApJ...616..485T}, but a higher extinction than would be estimated based on the line of sight to the object was adopted in order to justify such temperature.

\begin{floatingfigure}[lt]{9.5cm}
\includegraphics[scale=0.55,bb=88 200 580 620,clip=]{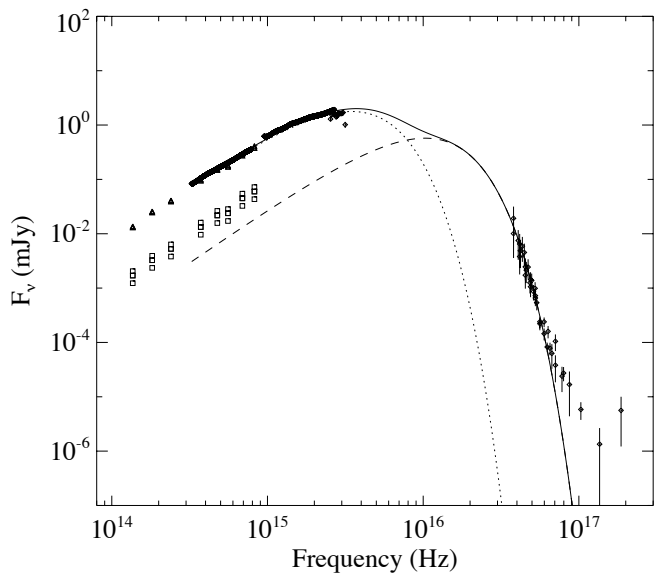}  
\caption{\footnotesize Spectral energy distribution of \sbs. Open circles  correspond to X-ray data. Filled and open diamonds  are optical \& UV data. The observed data are fitted with two BB presented with dotted \& dashed lines and their sum by a solid line.  The open triangles and open squares are SED of opt/UV and X-component respectively calculated by {\sl Nightfall} for a number of solutions. They all are adjusted to the flux of  opt/UV component. The discrepancy between {\sl Nightfall}  and  BB fits is due to the lack of consideration of irradiation, geometry and effects of limb-darkening etc., in BB fitting.}
\label{fig:1}   
\end{floatingfigure}

Subsequent optical photometry  \cite{2005AIPC..804..173N} showed that the most rational explanation of the light curve would require a Roche lobe filling compact star irradiated by a higher energy companion. It could explain the double-hump light curve with uneven depth of minima by ellipsoidal variation due to the changing aspect of the surface of the star twice during orbital period, with the side facing the hot  component being brighter than the opposite side. To resolve this issue, we performed X-ray observations using XMM-{\sl Newton} observatory.   The 25\,ksec  exposure revealed a supersoft X-ray source with a count rate of 0.035 c/s in the PN detector, 0.0029 c/s in the MOS1 detector, and 0.0048 c/s in the MOS2 detector. It appears that all useful photons from the source are distributed in a narrow region at the soft end of the sensitivity range of the telescope (100 -- 300 eV), which unfortunately suffers some calibration uncertainties \cite{2006ESASP.604..937S}, but the flux is very consistent in all three detectors. 

The spectral energy distribution (SED) of the object is presented in Fig.~\ref{fig:1}. The soft X-rays are certainly not produced by the component visible in the optical/UV range.  By adding the HST data to the SED, covering the near-UV where the extinction is better defined, we can state that the temperature of the optical/UV component is much lower than determined in \cite{2004ApJ...616..485T}.  In addition, there is another hotter component in the binary system.
A simple $\chi^2$ fitting of two black bodies (BB) permits to determine that the opt/UV component is about 60\,000\,K and the hot X-component is about 180\,000\,K. 

With rough knowledge of temperatures we can proceed and impose additional   constraints on the parameters of the system. One such constraint can be imposed considering the Population II origin of the object (another possibility includes the  capture of the object in a merger with a satellite galaxy) and dynamics of envelope expulsion by a post-AGB star. First of all, the core of a star heated above 55\,000\,K\footnote{The minimum temperature is defined not only by BB fitting, but also an absence of He\,{\small I} lines in the spectrum \cite{2004ApJ...616..485T}.}  in the post-AGB stage must be at least 0.56\msun\ \cite{1995A&A...299..755B}. But in a 3.9 hr Pop. II binary the $\ge 0.56$ \msun\  component undergoing ejection of envelope may descend from a ZAMS star of  $\approx 0.9-1.1$ \msun. 
Formation of a DD binary through a common envelope requires a certain combination of parameters of binding energy and CE efficiency  \cite{2007arXiv0704.0280W}, which, according to our calculations 
leads to the upper limit of the core mass of 0.58\msun.  Calculations were made similar to \cite{2007A&A...466.1031V}, using code described in \cite{2000MNRAS.315..543H}.

We also can place a firm lower limit on mass of the X-component. The deduced temperature  of the X-component certifies that it acts like a classic supersoft X-ray source (SSS) \cite{2006AdSpR..38.2836K}. It has been heated by a steady surface burning of matter accreted at high accretion rate during previous,  symbiotic stage of evolution. 
Only accretors with M$_{\rm wd} > 0.7$\msun\ reach temperatures high enough to be detected as SSS \cite{2001ASPC..229..309W}. They therefore occupy the upper range of WD masses (0.7-1.4) fed at a few$\times 10^{-7}$ \msun/year. 
Even the pessimistic estimate of the total accreted mass in this stage of evolution reaches 0.1-0.15 \msun \cite{2006MNRAS.372.1389L}, so even if the X-component was formed at the lower edge with M$_{\rm x}=0.7$\,\msun\ it easily may have grown to 0.83-0.85 \msun\ since then, which is enough for total mass of the system to reach Chandrasekar limit.

We can use these  estimates of temperatures  and masses of the binary components to simultaneously fit  radial velocity (RV)  and light curves (LCs) and calculate a set of important binary parameters using {\sl Nightfall}\footnote{ \url{http://www.hs.uni-hamburg.de/DE/Ins/Per/Wichmann/Nightfall.html}\\ Nightfall is based on a physical model that takes into account the nonspherical shape of stars in close binary systems, as well as mutual irradiance of both stars, and a number of additional physical effects.}. The program allows for fixing or setting as free parameters the mass ratio, inclination angle, filling factor of the corresponding Roche lobes, temperatures of components, total mass of the system, and component's separation. The albedo and limb darkening are also taken into account.  For input we used RV data described in  \cite{2004ApJ...616..485T} and LCs were used from Calar-Alto observations presented in \cite{2005AIPC..804..173N}. We calculated a grid of solutions with total mass fixed from 1.3 to 1.6\msun\ and did not find any significant difference in fits, though it looks like increasing total mass tends to improve them slightly. But more importantly, we got a handle on the sizes of  components, which are strongly bound to the shape of LCs. The size of the opt/UV component is confined to a narrow range of 0.4-0.48 \rsun\  as a consequence of fitting ellipsoidal light variation. On the other hand, the difference in depths of minima in LCs requires certain luminosity of X-component that can be achieved only if it is about 0.12-0.14\rsun. Attempts to squeeze the X-component size down to 0.05\rsun\ result in an unrealistically high temperature of X-component. The required temperature rises up to 250\,000\,K, which is not supported by blackbody fitting to the SED.  


\section{Nature of the object as type Ia SN progenitor}

\begin{floatingfigure}[lt]{9.2cm}
\includegraphics[scale=0.40,bb=5 0 760 440,clip=]{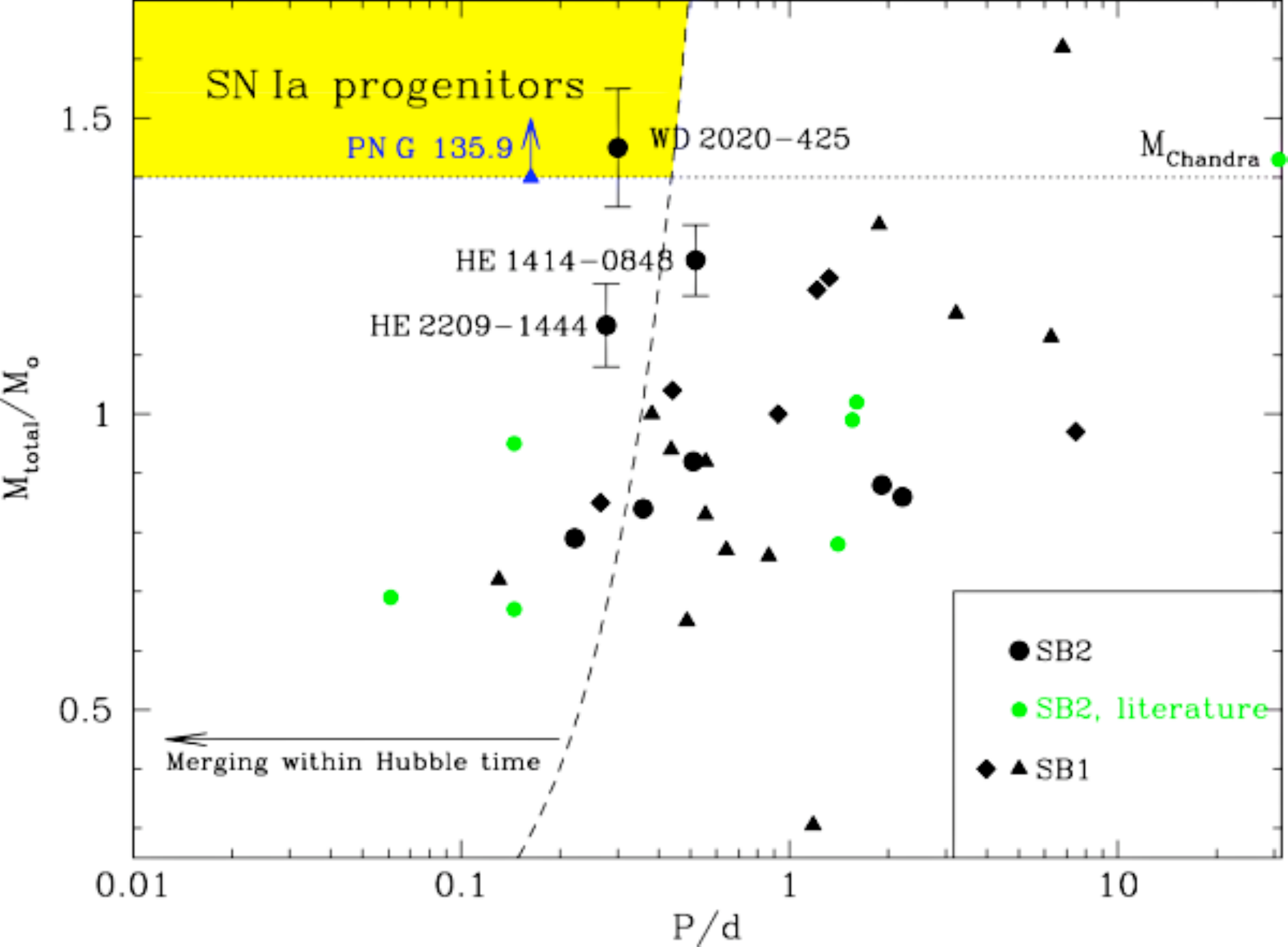}  
\caption{\footnotesize Total mass plotted against logarithmic period of DD and sdB+WD systems
from the SPY survey and literature. The \sbs=\pn\  is marked by an arrow.}
\label{fig:2}   
\end{floatingfigure}

We are observing the binary core of planetary nebula \sbs\  after expulsion of the common envelope by the component that currently descends from AGB to a WD (opt/UV component). It appears that the other component of the binary is also a WD heated by a residual burning of accreted material. The system follows closely a scenario described in detail in  \cite{2006MNRAS.372.1389L}. High-rate accretion has ceased probably not long ago, and we still see evidence of residual accretion judging from some excess of X-ray flux and flickering in the optical LC. Deduced radius of the X-ray component exceeds  that of an  ordinary WD, which we interpret as a hot atmosphere-like shell of accreted material around WD.  
The high temperature of the X-component and other facts point out that it underwent a period of very high rate accretion and leading us to a conclusion that it is a massive WD. The total of initial and accreted mass then may easily surpass 0.83\msun, which will make the system a likely SN Ia progenitor, given its short orbital period and possibility of its merging in a Hubble time (Fig.~\ref{fig:2}). That will complete the full evolutionary cycle.




\bibliographystyle{aipproc}   





\end{document}